\documentclass[twocolappendix, iop, apj]{emulateapj}
\usepackage{graphicx}
\usepackage{natbib}
\usepackage{amsmath}

\newcommand{\edit}[1]{#1}

\newcommand{\eg}{{\it e.g.,}} 
\newcommand{\be}{\begin{eqnarray}}
\newcommand{\ee}{\end{eqnarray}}
\def\HeI{\ion{He}{1}}
\def\HeII{\ion{He}{2}}
\def\HeIII{\ion{He}{3}}
\def\nne{\ensuremath{n_\mathrm{e}}}
\def\nijk{\ensuremath{n_{ijk}}}
\def\kB{\ensuremath{k_\mathrm{B} }}

\begin{document}
\title{Non-equilibrium helium ionization in an MHD simulation of the solar atmosphere}
\shorttitle{Non-equilibrium helium ionization}
\shortauthors{T. P. Golding, J. Leenaarts and M. Carlsson}

\author{Thomas Peter Golding}
 \affil{Institute of Theoretical Astrophysics, University of Oslo,
   P.O. Box 1029 Blindern, NO-0315 Oslo, Norway}
 \email{thomas.golding@astro.uio.no}
\author{Jorrit Leenaarts}
  \affil{Institute for Solar Physics, Department of Astronomy, Stockholm University, AlbaNova University Centre, SE-106 91 Stockholm, Sweden}
  \email{jorrit.leenaarts@astro.su.se}
\author{Mats Carlsson}
  \affil{Institute of Theoretical Astrophysics, University of Oslo,
    P.O. Box 1029 Blindern, NO-0315 Oslo, Norway}
  \email{mats.carlsson@astro.uio.no}

\begin{abstract}
\edit{The ionization state of the gas in the dynamic solar chromosphere can depart strongly from the instantaneous statistical equilibrium commonly assumed in numerical modeling.}
We improve on earlier simulations of the solar atmosphere that only included non-equilbrium hydrogen ionization  by performing a 2D radiation-magneto-hydrodynamics simulation featuring non-equilibrium ionization of both hydrogen and helium. The simulation includes the effect of hydrogen Lyman-$\alpha$ and the 
EUV radiation from the corona on the ionization and heating of the atmosphere. 
Details on code implementation are given.
We obtain helium ion fractions that are far from their equilibrium values. Comparison 
with models with LTE ionization shows that non-equilibrium helium ionization leads to 
higher temperatures in wave fronts and lower temperatures in the gas between shocks.
Assuming LTE ionization results in a thermostat-like behaviour with matter accumulating around the temperatures
where the LTE ionization fractions change rapidly. Comparison of
DEM curves computed from our models shows that non-equilibrium ionization leads to more radiating material in the temperature range 11-18\,kK compared to models with LTE helium ionization. We conclude that non-equilbrium helium ionization 
is important for the dynamics and thermal structure of the upper chromosphere and transition region. It
might also help resolve the problem that intensities of chromospheric lines computed from current models are smaller than those observed.
\end{abstract}


\section{Introduction}
The solar atmosphere is the Sun's dynamic outer layer, residing above the convection zone. 
Here, (partially) ionized gas interacts
with magnetic fields, and we see different kinds of waves, jets, and other phenomena. 
Numerical modeling is a powerful
tool for determining which physical processes are important in the solar 
atmosphere, either together with observations 
\citep[\eg][]{ortiz2014,depontieu_linebroadening2015},
or in their own right (where \cite{martinezsykora2012} and \cite{bourdin2015}
are examples).

Depending on the intended use, various levels of model sophistication 
are required. A model should include treatment of physics relevant for the 
regime under consideration.

We aim at modelling the solar atmosphere, all the way from the convection zone to the 
corona. One difficulty with this is the treatment of the chromosphere and 
transition region.
Here the gas goes from nearly neutral to almost completely ionized, the 
temperature increases from a few thousand kelvin to a million kelvin, and the 
dynamics goes from being dominated by the gas pressure to being dominated by 
the magnetic pressure. A reasonable description of the chromosphere and 
transition region is important for 
instance for coronal heating, a long standing problem in solar physics 
\citep{klimchuk2006,parnell2012},
since all the energy that ultimately ends up heating the corona, at some
point must have been transported through these regions.

In this paper we present a method for treating the time-dependent ionization
state of the atmosphere. Hydrogen is the most abundant element in the Sun, and its
non-equilibrium ionization has been found to result in higher temperature wave fronts
and lower temperature gas between wave fronts, compared to assuming local 
thermodynamic equilibrium 
\citep[LTE,][]{carlsson_stein2002,leenaarts2007}.
Although the number of helium particles is 
around ten times less than the number of hydrogen particles, its ionization state does have 
an effect on the energy balance of the upper chromosphere and transition region 
\citep{golding2014}. In fact, for a parcel of neutral solar gas, ionizing half of the helium atoms
requires an amount of energy equivalent to raising the temperature from 10\,kK to 20 kK.

An account of the non-equilibrium ionization state involves solving a set of rate equations
for the atomic population densities. The transition rate coefficients involve frequency 
integrals over the intensity, so a general description must necessarily take radiative 
transfer into account. 
The wave code presented in \citet{2003ApJ...589..988R} and
the code RADYN 
\citep{carlsson_stein1992, carlsson_stein1995, carlsson_stein1997} 
are examples of codes that include this complexity.
They solve the hydrodynamic equations 
and the rate equations together with the radiative transfer.
One serious drawback of these codes is that they operate only in one dimension. 
In 2D or 3D a detailed treatment of radiative transfer, such as that featured in the mentioned 
codes, is \edit{challenging} because the amount of computational 
work needed exceeds the capacity of current supercomputers. 

\cite{leenaarts2007} overcame this difficulty and used simplifying assumptions of the radiation 
field that enabled them to carry out 2D radiation-magnetohydrodynamics (RMHD) simulations, 
including the effects of non-equilibrium hydrogen ionization. These simulations were performed 
using the Oslo Stagger Code \citep{Hansteen2004, Hansteen+Carlsson+Gudiksen2007}, a predecessor of the stellar atmosphere code Bifrost \citep{gudiksen2011}.
This method was later implemented in the Bifrost code \citep{golding2010}. An 
example of a 3D simulation with this package has been made available for 
download\footnote{For download details see IRIS Technical Note 33 at 
https://iris.lmsal.com/documents.html}
\citep{2016A&A...585...A14}. We expand on this work by including 
also a description of non-equilibrium helium ionization. The paper is laid out in the following 
way: we explain the developed method in Section \ref{section:method}, and  present the 
results in Section \ref{section:results}. Finally we summarize and draw conclusions in Section 
\ref{section:conclusion}.

\section{Method}\label{section:method}

The stellar atmosphere code Bifrost solves the equations of 
radiation-magnetohydrodynamics (RMHD):
\begin{eqnarray}
  \frac{\partial \rho}{\partial t} &=& - \nabla \cdot (\rho \mathbf{u}) \label{eq:mass} \\
  \frac{\partial (\rho \mathbf{u})}{\partial t} &=& -\nabla \cdot (\rho \mathbf{u} 
                  \mathbf{u} - \mathsf{T}) - \nabla P + \mathbf{J} \times \mathbf{B} + \rho \mathbf{g} 
                  \label{eq:momentum} \\
  \frac{\partial e}{\partial t} &=& - \nabla \cdot (e \mathbf{u}) - P \nabla \cdot \mathbf{u} 
                  + Q_{\mathrm{r}} + Q_\mathrm{other} \label{eq:energy} \\
  \frac{\partial \mathbf{B}}{\partial t} &=& \nabla \times (\mathbf{u} \times \mathbf{B})
                      -\nabla \times \eta \mathbf{J} \label{eq:induction} \\
  \mu \mathbf{J} &=& \nabla \times \mathbf{B}, \label{eq:faraday}
\end{eqnarray}
where $\rho$ is the mass density, $\mathbf{u}$ is the velocity field, $\mathsf{T}$ is the stress
tensor, $P$ is the gas pressure, $\mathbf{J}$ is the current density, $\mathbf{B}$ is 
the magnetic field, 
$\mathbf{g}$ is the gravitational acceleration, $e$ is the internal energy density per 
unit volume, $Q_\mathrm{r}$ is the heating due to radiation, $Q_\mathrm{other}$ is the
heating due to \edit{heat conduction and viscous and ohmic dissipation}, 
$\eta$ is the magnetic diffusivity, and $\mu$ is
the vacuum permeability. We express radiative heating as the sum of 
different contributions, 
\begin{eqnarray}
  Q_{\mathrm{r}} = Q_{\mathrm{phot}} - L_\mathrm{chrom} + 
                              Q_\mathrm{Ly\alpha} + Q_\mathrm{EUV},
\end{eqnarray}
where $Q_\mathrm{phot}$ is the radiative heating from the photosphere described in 
\cite{gudiksen2011}, $L_\mathrm{chrom}$ is losses from the chromosphere due 
to strong lines, $Q_\mathrm{Ly\alpha}$ is heating from the 
Lyman-$\alpha$ line of hydrogen, and
$Q_\mathrm{EUV}$ is the heating from EUV photons, corresponding to the thin radiative losses
from the transition region and
corona (negative $Q_\mathrm{EUV}$) absorbed in the chromosphere (positive $Q_\mathrm{EUV}$). Recipes for the three latter
contributions are described in \cite{carlsson2012} (from now on CL12).
\defcitealias{carlsson2012}{CL12} These recipes are based on empirical fits.
In this work we discuss the effects 
of non-equilibrium ionization and the absorption of coronal radiation in the 
chromosphere. These effects enter the RMHD equations through $P$, 
$Q_\mathrm{Ly\alpha}$ and $Q_\mathrm{EUV}$. In the remaining part of
the method section we describe how we compute these quantities self-consistently
with the population densities of hydrogen and helium. For further details on the 
Bifrost code we refer the reader to \cite{gudiksen2011}.

\subsection{Equations of State}

To close the RMHD equation set, we need to relate $P$ to $\rho$ and $e$.  $P$
is in general given by the relation,
\begin{equation}
 P = \kB T (\nne + \sum_{i,j,k} \nijk), \label{eq:pressure}
\end{equation}
where $k_\mathrm{B}$, $T$, \nne\ and \nijk\ are the Boltzmann
constant, the gas temperature, the electron density, and the population density of an atom 
or molecule $i$ in the ionization stage $j$ occupying the excitation state $k$. 
The quantities $T$, \nne\ and \nijk\ are not present in the RMHD equations, and must be specified through extra equations. The MHD model requires a single temperature for all species, so we express  the internal energy density as
\begin{equation}
 e = \frac{3}{2} \kB T \left(\sum_{i,j,k} \nijk + \nne \right) + \sum_{i,j,k}  \nijk E_{ijk}  \label{eq:energyeq},
\end{equation}
where $E_{ijk}$ is the dissociation, ionization or excitation energy of an 
element or molecule $i$ in the ionization stage $j$ occupying the excitation 
state $k$. In addition, MHD assumes charge neutrality, so that the electron density is  given by
\begin{equation}
  n_\mathrm{e} = \sum_{i,j,k} (j-1) \nijk \label{eq:chargeeq},
\end{equation}
where $j=1$ denotes a neutral particle, $j=2$ a singly ionized particle, etc.
The total number of atomic nuclei of each element (including atoms bound in molecules)  is conserved, so we add conservation equations for each element. The additional equations required to constrain \nijk\ and close the set of equations 
depend on the physical system one wants to model.

We refer to the Equations \ref{eq:pressure}, \ref{eq:energyeq} and \ref{eq:chargeeq} and the extra equations used to constrain \nijk\ (see \ref{sec:LTEEOS} and \ref{sec:noneqabel}) collectively as the equations of state (EOS).

\subsubsection{The Local Thermodynamic Equilibrium EOS} \label{sec:LTEEOS}

The most common way of constraining the population densities is to assume local thermodynamic equilibrium 
(LTE). In that case the temperature, electron density, and population densities are related through a set of Saha-Boltzmann equations and molecular equilibrium equations
\citep[\eg][]{mihalas1978}.
This combination of equations leads to an EOS that is local: once the internal energy, mass density, and the elemental abundances are given, $T$, \nne, and \nijk, and hence $P$, can be computed directly. Based on these variables other relevant quantities, such as opacities, can be computed. This method is computationally fast because all relevant variables can be pre-computed and read from a table. Most stellar atmosphere codes use this assumption
\citep[\eg][]{1982A&A...107....1N, 2005A&A...429..335V,gudiksen2011, 2012JCoPh.231..919F,2015arXiv150707999W}.
The Bifrost implementation employs tables that depend on $\rho$ and $e$.

\subsubsection{Non-Equilibrium Ionization of Abundant Elements} \label{sec:noneqabel}

The assumption of LTE breaks down in the layers above the solar photosphere. Radiative transition 
rates become dominant over collisional rates leading to non-local coupling of different parts of the 
atmosphere through radiation. The transition rates themselves become so small that the ionization-
recombination and molecular association-dissociation timescale can become long compared to 
typical hydrodynamical timescales in the atmosphere
 \citep{1979ApJS...40..793J}.

In that case the population densities should be determined from a continuity equation:
\begin{equation}
 \frac{\partial \nijk}{\partial t} + \nabla \cdot (\nijk \boldsymbol{u}) = \mathrm{gains}-\mathrm{losses}
                   \label{eq:rateeq}.
\end{equation} 
The gains and losses represent processes that add or remove particles out of state $ijk$, for example collisions with electrons, radiative transitions, or processes that form or destroy molecules. This type of lack of any equilibrium in the population densities is generally referred to as {\it non-equilibrium}.

The terms that matter the most in the EOS are those that are associated
with the most abundant elements. In the solar atmosphere these are hydrogen and helium 
\citep{2009ARA&A..47..481A},
which are both susceptible to non-equilbrium effects:
Non-equilibrium hydrogen ionization leads to increased temperature fluctuations throughout the chromosphere as shown in detailed 1D wave simulations of the solar atmosphere
\citep{carlsson_stein2002}.
\cite{leenaarts2011}
investigated non-equilibrium formation of H$_2$  molecules, which can lead to very low temperatures in the chromosphere because the exothermic H$_2$ formation rate is too low to counteract the rapid adiabatic cooling between internetwork shocks.
\cite{golding2014}
showed that non-equilibrium helium ionization leads to significant differences in temperature in the upper chromosphere and transition region.

Solving the continuity equations involves solving the complete radiative transfer problem because 
radiative rates enter into the gain and loss terms in Equation~\ref{eq:rateeq}. Each radiative
transition rate coefficient is found by integrating the mean intensity over frequency. 
The mean intensity for each frequency is essentially found by integrating the source function over the 
computational domain. This procedure is too costly to apply directly in a multi-dimensional RMHD 
code where it must be repeated every time step, and typical time steps are of the order of 
milliseconds. Simplifications that lead to a computationally tractable problem are fortunately possible.

\subsection{Non-Equilibrium Hydrogen in Bifrost}\label{section:oldhion}

The Bifrost code already includes an option to compute the EOS including non-equilibrium hydrogen ionization and H$_2$ molecule formation using such simplifications. A detailed description of the method can be found
in \cite{leenaarts2007,leenaarts2011} and \cite{gudiksen2011}. Here we briefly restate the essential points as a foundation for the further extensions to the method that are the topic of this paper.

The non-equilibrium hydrogen ionization method is based on approximations by
\cite{Sollum1999}
who observed that the mean intensity in non-Lyman hydrogen transitions decouples from the gas temperature in the
photosphere, and that the mean intensity above this decoupling point stays constant.
This means that each radiative rate coefficient in the chromosphere and above can be described by one parameter - the
radiation temperature above the photosphere. In the photosphere and below the radiation temperature is simply the gas temperature. 

The continuity equations for atomic hydrogen read
\begin{equation}
 \frac{\partial n_i}{\partial t} + \nabla \cdot (n_i \boldsymbol{u}) = \sum_{i \ne j} n_j P_{ji} - n_i \sum_{i \ne j} P_{ij},
     \label{eq:rateeq_hion}
 \end{equation} 
where $P_{ij}=C_{ij}+R_{ij}$ is the sum of the collisional ($C_{ij}$) and radiative ($R_{ij}$) rate coefficients and $n_i$ are population densities in the  excitation and ionization stages of hydrogen. The collisional rate coefficients depend on the local electron density and temperature only, and the non-Lyman radiative rate coefficients are computed from Sollum's radiation temperatures, so that the equations are again local. For all other elements we assume LTE.

The assumption that the radiation field decouples from the temperature in the photosphere does not hold for the Lyman transitions.  Instead, these are assumed to be in detailed  radiative balance, i.e. $n_1R_{1j} = n_jR_{j1}$, which is equivalent to
\edit{setting $R_{1j}=R_{j1}=0$ in the hydrogen continuity equations} (where subscript 1 denotes the ground state). 

A continuity equation for H$_2$ is also included, with three-body association and collisional dissociation with neutral hydrogen atoms as source and sink terms.

The continuity equations are solved using operator splitting. First the populations are advected using an explicit first order upwind scheme. The source and sink part are solved implicitly together with Equations~\ref{eq:energyeq} and~\ref{eq:chargeeq}, for the population densities, electron density and temperature. The pressure then follows from Equation~\ref{eq:pressure}.

We now describe the new additions to Bifrost: the inclusion of hydrogen Lyman transitions in Section~\ref{sec:lyman} and non-equilibrium helium ionization in Section~\ref{section:method_helium}.

\subsection{Heating and cooling in Lyman-$\alpha$} \label{sec:lyman}

The assumption that the Lyman transitions are in detailed balance breaks down in the upper chromosphere 
\citep[see for example][]{1981ApJS...45..635V},
and the method described in Section~\ref{section:oldhion} will thus not be accurate there.

Lyman-$\alpha$ photons are, roughly speaking, released from the 
transition region or from shocks, and absorbed by surrounding
cold chromospheric material 
\citepalias{carlsson2012}.
The upward rate coefficient for a bound-bound transition depends
on the radiation field. Computing the frequency-dependent radiation field every time we advance the RMHD equations with a timestep is computationally too expensive. Therefore we use a simple one-frequency
approach. Lyman-$\alpha$ photons that contribute 
to the transport of energy are those that manage to escape from the 
location where they are emitted. Deep in the solar atmosphere most 
Lyman-$\alpha$ photons are emitted and absorbed at the same location, i.e. the
radiation temperature is equal to the gas temperature. These
photons do not affect the energy balance of the gas, so we do not model them. 
We only consider the photons that are able to escape from where they are emitted.

\citetalias{carlsson2012} use an escape probability to parameterize hydrogen losses.
We adopt this escape probability to account only for the photons that make a difference in the
energy budget. The escape probability at a specific location, $E$, is modelled as 
a function of the column density of neutral hydrogen ($\tau$). This escape probability is 
monotonically decreasing with increasing $\tau$. In the corona and transition region
there is very little neutral hydrogen in the column above, so $E = 1$. Further down in the atmosphere $\tau$ increases and the escape probability goes 
to zero. We let the  downward radiative rate coefficient in the Lyman-$\alpha$ transition be proportional to the escape probability,
\begin{eqnarray}
  R_{21} &=& A_{21}E(\tau), \label{eq:rjilya}
\end{eqnarray}
where $A_{21}$ is the Einstein coefficient for spontaneous deexcitation.
From this rate coefficient we express the frequency-integrated Lyman-$\alpha$ emissivity,
\begin{eqnarray}
  \eta_{\mathrm{L}\alpha} &=& \frac{h\nu_0}{4\pi} n_2R_{21} \label{eq:etalya},
\end{eqnarray}
where $h$ is Planck's constant, $\nu_0$ is the line center frequency, and $n_2$ is the
number density of the upper level of the line.
We ignore stimulated emission and express the  Lyman-$\alpha$  opacity as
\begin{eqnarray}
 \chi_{\mathrm{Ly}\alpha} = \frac{h\nu_0}{4\pi}n_1 B_{12} \phi,\label{eq:chilya}
\end{eqnarray}
where $B_{12}$ is the Einstein coefficient for radiative excitation and $\phi$ is a frequency-averaged profile
function. We set $\phi = 2 \times 10^{-12}$ Hz$^{-1}$ which is half of the maximum value 
of the corresponding frequency-dependent Doppler profile at 10 kK. 
From the emissivity and opacity we obtain the frequency-integrated mean intensity,
$J_{\mathrm{Ly}\alpha}$,  
by solving the equation of radiative transfer
using a short-characteristics method for decomposed domains
\citep{hayek2010,carlsson2012}.
The upward radiative rate coefficient is then expressed as
\begin{eqnarray}
  R_{12} = B_{12} J_{\mathrm{Ly}\alpha} \label{eq:rijlya}.
\end{eqnarray}

When these non-zero radiative rate coefficients in the Lyman-$\alpha$ transition 
are included in the hydrogen rate equations, we compute the Lyman-$\alpha$ heating,
\begin{eqnarray}
 Q_{\mathrm{Ly}\alpha}  = h\nu_0(n_1R_{12} - n_2R_{21}), \label{eq:qlya}
\end{eqnarray}
where $n_1$ and $n_2$ denote the population densities of the lower and upper level. 
All other Lyman lines are still assumed to be in detailed balance. The Lyman continuum is however taken into account, as described in Section~\ref{section:method_helium}.

\edit{Our handling of the Lyman-$\alpha$ transfer represents an extreme simplification. 
Nevertheless, we find it worthwhile. Our description qualitatively produces what we want: 
cooling in the transition region and in wave fronts, and heating in the colder ambient gas.}

\subsection{Non-Equilibrium Helium Ionization}\label{section:method_helium}

\begin{figure}
 \includegraphics[width=\columnwidth]{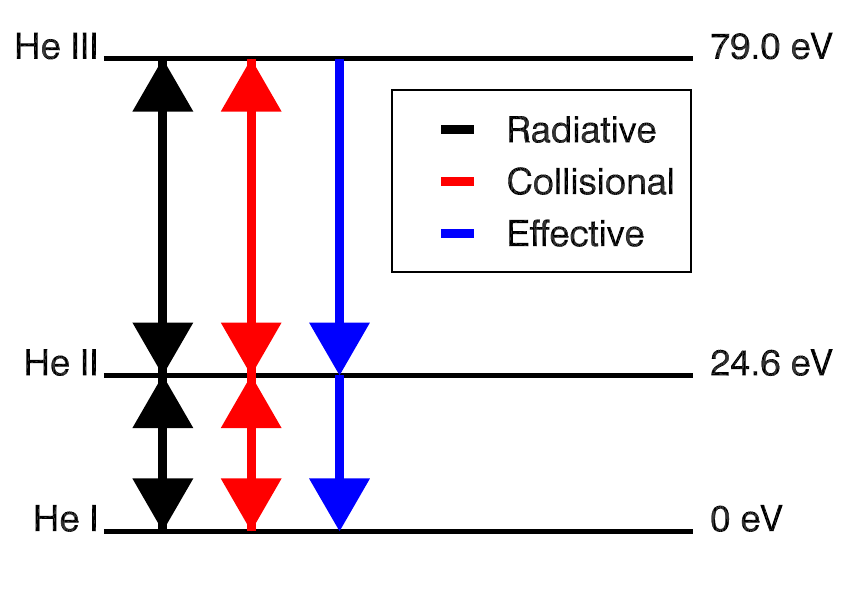}
 \caption{Simplified helium 3 level model atom. It consists of the
                  ground states of each ion stage. Transitions taken into account are 
                 photoionization, radiative recombination, collisional
                 ionization by electrons, three body recombination, and
                 an effective recombination modelling recombination
                 to excited states and the subsequent radiative cascade to the ground state.}
 \label{fig:fig_model_he3}
\end{figure}

We obtain non-equilibrium helium population densities by solving the rate
equations for helium (Eq. \ref{eq:rateeq}). We use a three level model atom,
consisting of
the ground states of  \HeI\ and \HeII, and the doubly ionized state 
\HeIII\ (see Fig. \ref{fig:fig_model_he3}). 
\citet{golding2014} 
showed that the model atom reproduces the correct ionization state remarkably well given
its extreme simplification.

The model atom includes collisional ionization/recombination 
\citep[][see Section \ref{section:coll_coeff}]{ar1985},
photoionization/recombination
and an extra recombination rate coefficient that mimics the net radiative recombination
to excited states. Further details and the derivation of the extra recombination rate coefficient 
are given in 
\cite{golding2014}. 
The photoionization rate coefficient 
is a frequency integral over the mean intensity weighted by the photoionization 
cross section. As mentioned earlier, this is a quantity which is too computationally 
costly to be computed in detail. To make the problem computationally tractable we use a bin 
formulation, reducing the integral to a sum with only a few terms.
It is the EUV photons emitted downward from the transition region and corona 
that are responsible for ionizing helium in the chromosphere. Photons 
with wavelengths shorter  than 504 \AA\ are prone to ionize neutral helium. However, 
they might just as well ionize hydrogen in the Lyman continuum. The Lyman continuum
has its ionization edge at 911 \AA. We include the hydrogen Lyman continuum transition and
for that reason we choose the first bin to have its upper limit at 911 \AA. 
We let the last bin have a lower limit at wavelength 20 \AA, 
well below the photoionization edge wavelength for \ion{He}{2} at 228 \AA. 

\begin{figure}
 \includegraphics[width=\columnwidth]{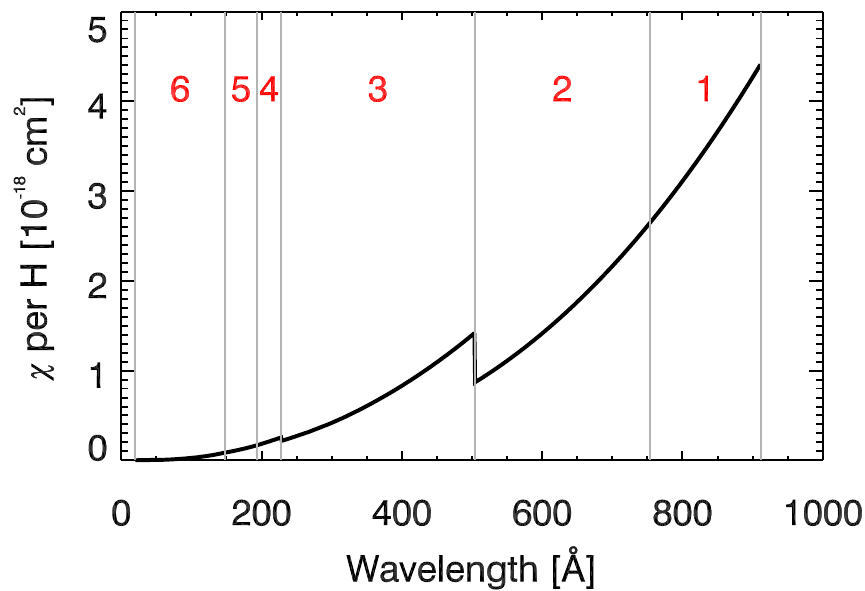}
 \caption{Opacity per hydrogen atom in typical upper chromospheric
                 material.
                 The red numbers 
                 name the bins used for the simulations presented in this paper.}
 \label{fig:fig_euv_bins}
\end{figure}

Figure \ref{fig:fig_euv_bins} shows the continuum opacity per hydrogen 
atom as a function of wavelength for a parcel of solar gas where hydrogen is
30\% ionized, helium is 75\% neutral and 25\% singly ionized -- values
typical for the upper chromosphere. Here, and in the transition region, is where 
the ionization state of helium is most important for the energy balance.  
Radiation bins are indicated in the figure. 
They are chosen in the following way:
bin boundaries are set at the photoinioniziation edges of \HeI\ and
\HeII, resulting in three main bins which are further split into a number of sub-bins.

In each of the main bins the continuum opacity falls off with decreasing wavelength.
We set the boundary between sub-bins at the wavelengths where the relative 
change of opacity is equal in each of the sub-bins.
The upper and lower frequency of the bin $j$ is denoted $\nu_{j,0}$ and $\nu_{j,1}$.
We adopt a formulation where x denotes the transition and $j$ is the bin number.
$n_{\mathrm{g}, \mathrm{x}}$ and $n_{\mathrm{c},\mathrm{x}}$ are the population
densities of the ground and continuum states. 

There are three continuum transitions, the ground state \ion{He}{1} continuum, 
the ground state \ion{He}{2} continuum, and 
hydrogen Lyman continuum. That means that we in
principle need 3$N_\mathrm{bin}$ photoionization cross sections ($N_\mathrm{bin}$ 
denotes the total number of bins). The photoionization
cross section for transition x in bin $j$ is denoted $\sigma_{\mathrm{x},j}$. Some of 
these photoionization cross sections are zero. For instance, $\sigma_{\mathrm{x},j}$ 
corresponding to the transition between ground states \ion{He}{1} and \ion{He}{2} is 
zero in all bins that have their lower wavelength boundary at 504 \AA \ or higher.   

We find $\sigma_{\mathrm{x},j}$ for a transition x in a specific bin $j$ by
equating the upward radiative rate coefficients in the binned and continuum 
formulation,
\begin{eqnarray}
  \sigma_{\mathrm{x},j} = \frac{4\pi\int_{\nu_{j,0}}^{\nu_{j,1}} \frac{J_\nu}{h\nu} 
           \sigma_{\mathrm{x},\nu}\ d\nu}
                         {4\pi J_{j} \int_{\nu_{j,0}}^{\nu_{j,1}} \frac{1}{h\nu}\ d\nu} 
                         \label{eq:photoxbin} 
\end{eqnarray}
where $J_\nu$ and $\sigma_{\mathrm{x},\nu}$ are the the frequency dependent mean intensity
and the frequency dependent photoionization cross section for the transition x. $J_j$ is the 
mean intensity that is found using $\sigma_{\mathrm{x},j}$ in the opacity. This equation is 
solved by iteration for a point in the chromosphere of a reference atmosphere. 
As reference atmosphere we use the initial snapshot of the simulations
from
 \cite{golding2014}.
Having all the atomic constants determined, we can 
compute the photoionization and radiative recombination rate coefficients,
\begin{eqnarray}
  R^{\mathrm{up}}_{\mathrm{x}} &=& \sum_{j=1}^{N_\mathrm{bin}} 
                                      W_{\mathrm{x},j} J_j \label{eq:rij} \\ 
  R^{\mathrm{down}}_{\mathrm{x}} &=& \left[\frac{n_{\mathrm{g},\mathrm{x}}}
        {n_{\mathrm{c},\mathrm{x}}}\right]_\mathrm{LTE}\sum_{j=1}^{N_\mathrm{bin}} 
                                      W_{\mathrm{x},j} B_j \label{eq:rji},
\end{eqnarray}
where $W_{\mathrm{x},j}$ is a bin-dependent constant, $[n_{\mathrm{g},
\mathrm{x}}/n_{\mathrm{c},\mathrm{x}}]_\mathrm{LTE}$ is the LTE ground 
state to continuum population density ratio for the transition x, given by the 
Saha relation 
\citep[see for instance][]{mihalas1978}, 
and $B_j$ is the mean
of the Planck function corrected for simulated emission. We ignore stimulated 
emission, so that the radiative recombination coefficient is independent of the 
mean intensity. A detailed derivation of these
expressions is given in Section \ref{section:rate_coeff}.

Test computations showed that the bin divisions drawn in Figure 
\ref{fig:fig_euv_bins} reproduce the actual photoionization rate coefficients of
the reference atmosphere fairly well. These six bins are the ones we use in the 
simulations presented in this paper.

\subsubsection{EUV Radiative Transfer}\label{section:euv_rad}

\begin{figure}
 \includegraphics[width=\columnwidth]{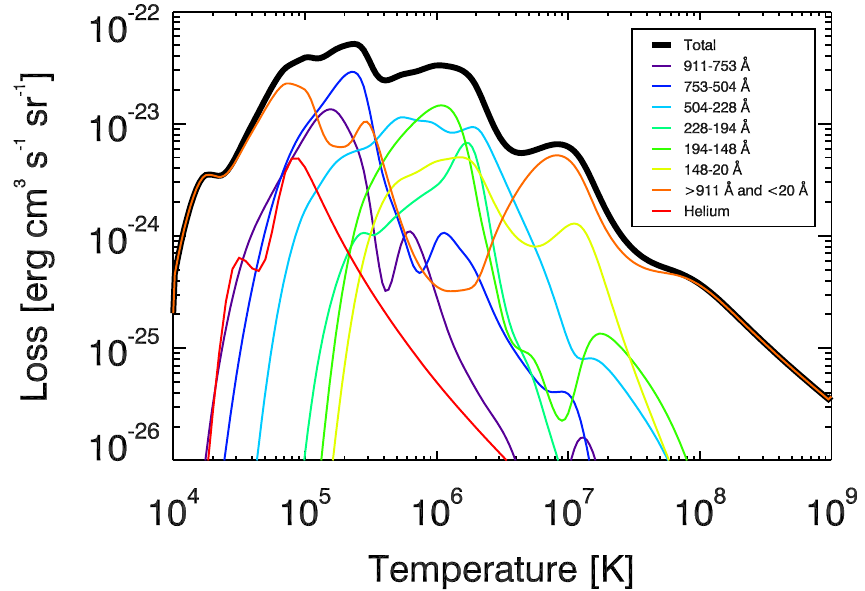}
 \caption{Coronal radiative losses for the 6 bins used in our simulations. These losses
 	        do not include contributions from hydrogen and helium. The 
	        helium losses that are used in the simulations
	       without non-equilibrium helium ionization are shown in red.}
 \label{fig:fig_binned_cemiss}
\end{figure}

We compute the heating and cooling in  the EUV spectrum based on
the actual population densities of H and He. The bin opacity and bin emissivity
are necessary to compute the mean intensity, $J_j$. The opacity contribution for transition
x in the $j$th bin is  
\begin{eqnarray}
  \chi_{\mathrm{x},j} = n_{\mathrm{g},\mathrm{x}} \sigma_{\mathrm{x},j},
  \label{eq:opacity}
\end{eqnarray}
where $n_{\mathrm{g},\mathrm{x}}$ is the lower level population density of the transition.
We include two contributors of photons to the radiation. First, the photons released
when hydrogen and helium ions recombine. They give rise to an emissivity that is 
transition and bin dependent, 
\begin{eqnarray}
  \eta_{\mathrm{x},j} = n_{\mathrm{c},\mathrm{x}} \left[ 
  \frac{n_{\mathrm{g},\mathrm{x}}}{n_{\mathrm{c},\mathrm{x}}} \right]_{\mathrm{LTE}}
  \sigma_{\mathrm{x},j} B_j,
\end{eqnarray}
where $n_{\mathrm{c},\mathrm{x}}$ is the upper level population density of the transition.

Then, second, we include 
photons produced by collisional excitation followed by radiative deexcitation in lines in the transition region and corona. An exact account of the resulting emissivity 
is not possible, as it would require of us to solve the rate equations for all relevant
ions. Rather, we assume ionization equilibrium and sum up the line losses from relevant
ions in each radiation bin. Photons from 
lines with wavelengths larger than 911 \AA \ or shorter that 20 \AA \ are treated
as photons which, after emission, do not interact with matter, i.e. we have no opacity for these 
wavelengths. We can express this coronal emissivity in the $j$th bin as 
\begin{eqnarray}
  \eta_{\mathrm{cor},j} = \frac{n_{\mathrm{e}}n_\mathrm{H} L_{j}(T)}{\nu_{j,1}-\nu_{j,0}}
  e^{-P/P_0},
  \label{eq:cooling}
\end{eqnarray}
where $L_j(T)$ is a temperature-dependent loss function per electron per hydrogen
atom per steradian. The exponential factor ensures that we avoid including losses from 
hot regions that are dense. We set $P_0$ to a value typical
of the upper chromosphere to make sure these losses only come from
the transition region and corona.  We use \hbox{CHIANTI} to compute 
$L_j(T)$ with the abundances given in \texttt{sun\_coronal.abund} and the
ionization equilibrium values from \texttt{chianti.ioneq} 
%
\citep{dere1997, dere2009}. 
Figure \ref{fig:fig_binned_cemiss} shows $L_j(T)$ for the bins we use in the simulations
presented here. Losses due to helium are dominated by the \HeII\ 304 \AA\ line, and therefore we
include helium line losses using the non-equilibrium \HeII\ population:
\begin{eqnarray}
  L_{304}(T,n_{\mathrm{HeII}}) =  \left(\frac{h\nu_0 }{4\pi}\right)
  \left(\frac{n_{\mathrm{{HeII}}}}{n_\mathrm{H}}\right) q(T). \label{eq:he304}
\end{eqnarray}
Here the first factor is the photon energy per steradian, the middle factor
is the number density of singly ionized helium per hydrogen atom and 
$q(T)$ is the collisional excitation rate coefficient per electron taken  from CHIANTI. $L_{304}$ is added to the appropriate 
$L_j$. The total bin emissivity is then finally expressed as the sum of both 
the contributions,
\begin{eqnarray}
  \eta_j = \sum_{\mathrm{x}} \eta_{\mathrm{x},j} + \eta_{\mathrm{cor},j}.
\end{eqnarray}
We obtain the bin mean intensity, $J_j$, with the same formal solver that is
used for the Lyman-$\alpha$ radiation.

We now have all relevant quantities and can compute the heating rate in the 
upper chromosphere, transition region and corona 
caused by the EUV spectrum:
\begin{eqnarray}
  Q_{\mathrm{EUV}} =  4\pi \sum_{j=1}^{N_{\mathrm{bin}}} (\chi_j J_j - \eta_j )
  (\nu_{j,1}-\nu_{j,0}). \label{eq:qeuv}
\end{eqnarray}

\subsection{Four Simulation Runs}

\edit{In order to compare the various approximations to the ionization 
balance of hydrogen and helium we perform a differential study. Therefore 
we performed four two-dimensional simulation runs starting out from the same 
initial snapshot. The simulations are meant to be comparable to quiet sun conditions.
The spatial} domain spans a region
15.8 Mm $\times$ 16.6 Mm with an equidistant horizontal resolution of 33 km 
and a vertical resolution of 28 km at $z<5$ Mm, continuously increasing to 150 km
in the corona at $z>9$ Mm. $z=0$ Mm is set in the photosphere where 
the optical depth at 5000 \AA ~ is unity. The initial magnetic field has a mean absolute
value of 65 G at $z=0$ Mm. The flux is concentrated in four regions separated roughly 
by 4 Mm, the strongest of which has a negative sign ($x=10$\,Mm) and the three remaining 
concentrations slightly weaker and with a positive sign. 
\edit{The two concentrations at $x=2$\,Mm and $x=10$\,Mm form footpoints for 
looplike structures, and we refer to them as network.
In 2D models there is not enough magnetic dissipation to  sustain coronal 
temperatures self consistently. We therefore}
use a hotplate boundary condition in the corona that will heat or cool the plasma
towards a temperature of 1 MK on a timescale of around 400 seconds.
The runs differ in the way the EOS, and thus also $Q_\mathrm{Ly\alpha}$ and
$Q_\mathrm{EUV}$, is modeled.

The first run (LTE) treats all elements, including H and He, in LTE. $Q_{\mathrm{Ly\alpha}}$ 
and $Q_{\mathrm{EUV}}$ are computed from the recipes in
 \citetalias{carlsson2012}.

The second run (HION) treats hydrogen in non-equilibrium as described in \ref{section:oldhion}, with the Lyman transitions in detailed balance. All other elements are in LTE.  $Q_{\mathrm{Ly\alpha}}$ and $Q_{\mathrm{EUV}}$ are computed from the recipes given in 
 \citetalias{carlsson2012}.

The third run (LYA-HION) treats hydrogen in non-equilibrium as described in Section~\ref{section:oldhion}, but with the Lyman-$\alpha$ transition computed using our one-frequency recipe (see Section~\ref{sec:lyman}). $Q_\mathrm{Ly\alpha}$ is computed as described in Equation \ref{eq:qlya} and $Q_\mathrm{EUV}$ is computed from the recipe in
 \citetalias{carlsson2012}.

Finally, the fourth run (HELIUM) treats both hydrogen and helium in non-equilibrium. $Q_{\mathrm{Ly\alpha}}$ is computed from Equation \ref{eq:qlya} and $Q_\mathrm{EUV}$ is computed from Equation \ref{eq:qeuv}.

All the runs include tabulated losses from helium in the coronal emissivity (see Figure 
\ref{fig:fig_binned_cemiss}), except the HELIUM-run which includes helium losses as 
given by Equation \ref{eq:he304}. 

Compared to running with an LTE EOS, the computing time needed per time step is  
2-3 times longer when using a non-equilibrium hydrogen EOS and 4-5 times longer when
using the non-equilibrium hydrogen and helium EOS. The four simulations are run for at least 3000 
solar seconds and snapshots are written every 10 seconds. 
The details of each run are summarized in Table~ \ref{table:simulations}.

\begin{deluxetable}{rrrrrr}
\tablecaption{Simulation overview}
\tablenum{1}
\tablehead{\colhead{Simulation} & \colhead{H} & \colhead{ $Q_\mathrm{Ly\alpha}$} & \colhead{He} & \colhead{He304} & \colhead{$Q_\mathrm{EUV}$} \\ 
\colhead{} & \colhead{} & \colhead{} & \colhead{} & \colhead{} & \colhead{} } 
\startdata
LTE & LTE & recipe & LTE & eq. & recipe\\
HION & non-eq. & recipe & LTE & eq. & recipe \\
LYA-HION & non-eq. & detailed & LTE & eq. & recipe\\
HELIUM & non-eq. & detailed & non-eq. & non-eq. & non-eq. \\
\enddata
\label{table:simulations}
\end{deluxetable}

\section{Results}\label{section:results}
\begin{figure*}
 \includegraphics[width=\textwidth]{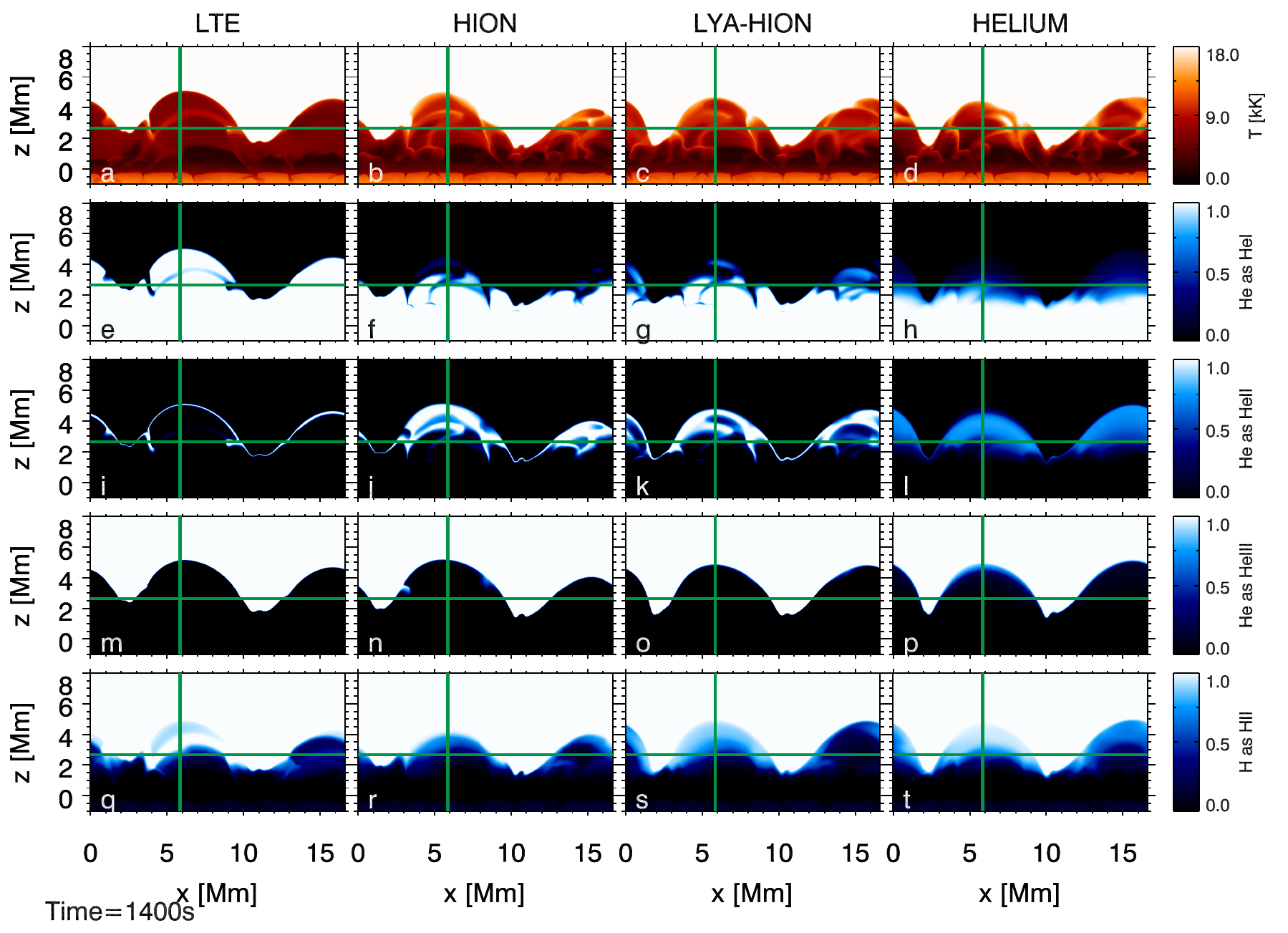}
   \caption{State of the atmosphere in the four different simulation runs at 1400~s 
   of solar time. Each column correspond to a given simulation. 
   The rows show, from top to bottom: temperature, fraction of He as \ion{He}{1}, fraction
   of He as \ion{He}{2}, fraction of He as \ion{He}{3}, and fraction of H as \ion{H}{2}. 
   Including non-equilibrium ionization leads to more structure in temperature and less 
   structure in the various ion fractions. This is because the finite transition rates limits how
   fast the ionization state can change. Changes in the internal energy will 
   manifest themselves as changes in temperature. Hydrogen is the most abundant element 
   and its non-equilibrium description has the strongest effect, as is seen by comparing 
   panels (a) and (b). Including also a non-equilibrium description of helium leads to a larger 
   temperature difference between the shock fronts and the plasma between the shocks than
   what we see in the LYA-HION run (compare panels (c) and (d)). The two green lines indicate 
   the cuts used for the space-time diagrams shown in Figures~\ref{fig:zt_tempion} and \ref{fig:xt_temperature}.}
 \label{fig:tg_ions}
\end{figure*}
\begin{figure*}
 \includegraphics[width=\textwidth]{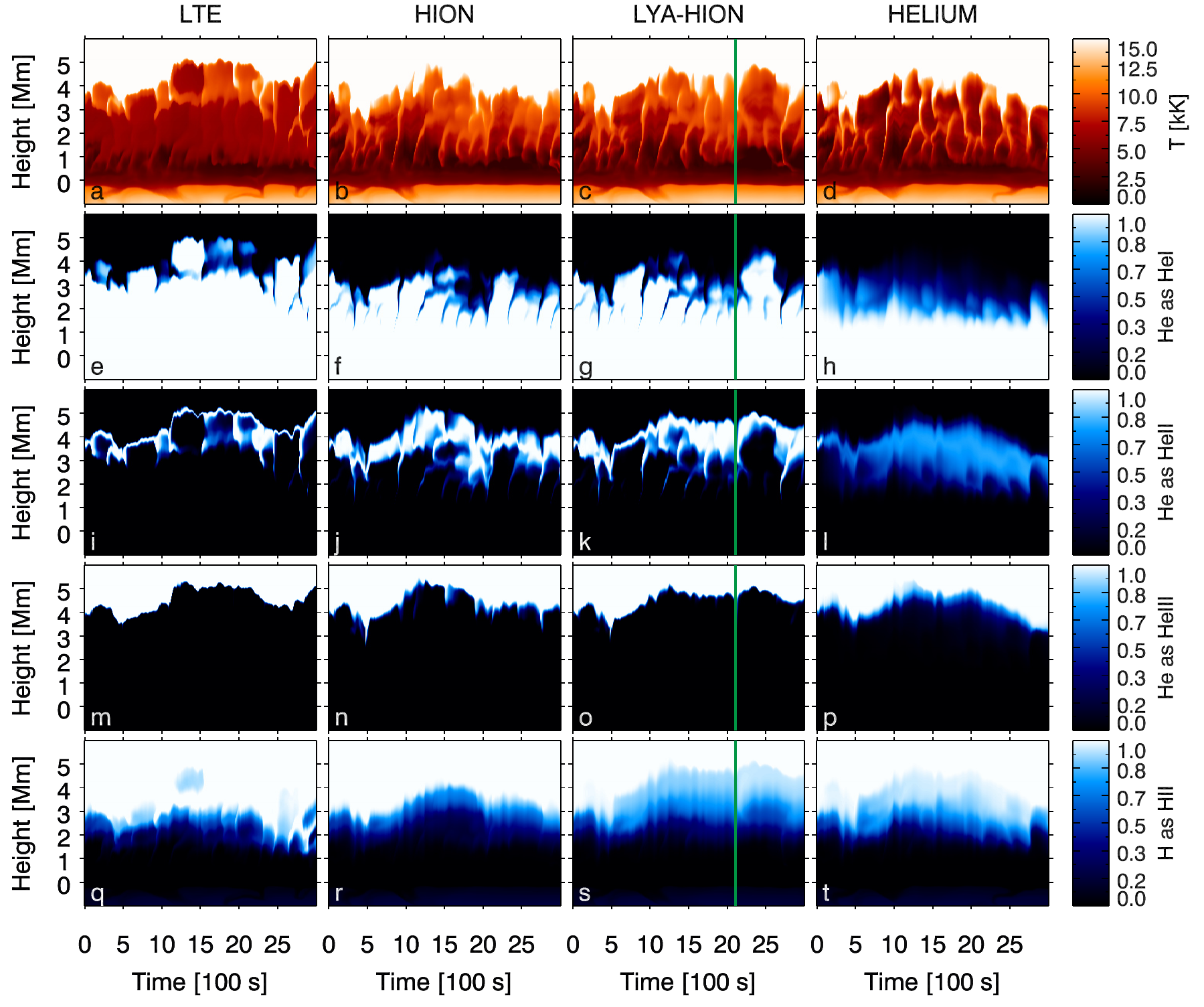}
   \caption{Evolution in temperature and ion fractions along a vertical cut in the four simulation
   runs. The cut used is indicated by the green vertical line drawn in all 
   the panels of Figure~\ref{fig:tg_ions}. 
   The columns, from left to right, are the different simulation runs. The rows show from top
   to bottom: temperature, fraction of He as \ion{He}{1}, fraction of He as \ion{He}{2}, 
   fraction of He as \ion{He}{3}, and fraction of H as \ion{H}{2}.  The LTE temperature in the
   chromosphere lies around two preferred values, 6 and 10 kK. When 
   non-equilibrium hydrogen is included (the three other runs), 6 kK no longer is a temperature
   at which the chromospheric gas stabilizes. The preferred value of 10 kK is not
   present in the HELIUM run. These effects can be explained by the ionization fractions: the larger
   variations in the temperature are compensated by less variations in the ion fractions.
   The green vertical line in the LYA-HION run panels indicates the column used in Figure 
   \ref{fig:lya}.}
 \label{fig:zt_tempion}
\end{figure*}

\subsection{Effect of Non-Equilibrium Hydrogen Ionization}\label{section:hydrogen}

Figure \ref{fig:tg_ions} shows temperatures and ion fractions at $t=1400$ seconds for all four
simulation runs.
We first focus our attention on the two left columns, corresponding to the 
LTE-run and the HION-run. Comparing panels (a) and (b), we see that the HION-run has a
hotter chromosphere and more contrasted structure in the temperature than the LTE-run. 
This is due to long hydrogen ionization recombination timescales, resulting in a slowly changing
ionization state. The fraction of H as \ion{H}{2} in the chromosphere of the
HION-run (panel (r)) is more stable and less responsive to waves (and other 
perturbations) than the ion fraction from the LTE-run (panel (q)). 
The change of internal energy associated with the waves will either go into ionizing atoms
(i.e. bound in ions) or increase the temperature.
We see this clearer in the vertical space-time cut shown in Figure~\ref{fig:zt_tempion}.  
The location of the cut is indicated by the green vertical lines of Figure~\ref{fig:tg_ions}. 
We can see waves propagating in the chromosphere in panels (a) and (b) as the tilted brighter
lines. The amplitude in temperature of the wave fronts is higher for the HION-run than for the 
LTE-run. Again, the hydrogen ion fraction of the HION-run (panel (r)) is less structured than the 
LTE-run ion fraction (panel (q)). This confirms what was reported in \cite{leenaarts2007}. 

Helium is in LTE in both the LTE-run and in the HION-run. Panels (i) and (j) in Figures~\ref{fig:tg_ions}
and \ref{fig:zt_tempion} show the fraction of He as \ion{He}{2} for the two simulations. For the 
region below the arched structures separating the chromosphere from the corona, this fraction
is larger in the HION-run than in the LTE-run. This is simply because of
the higher chromospheric temperatures featured in the HION-run. Neutral helium ionizes at ~10~kK in  LTE. The two ion fractions are different in value, but they
correlate well with the patterns and structures seen in temperature.

\subsubsection{Lyman-$\alpha$ Heating}

\begin{figure}
 \includegraphics[width=\columnwidth]{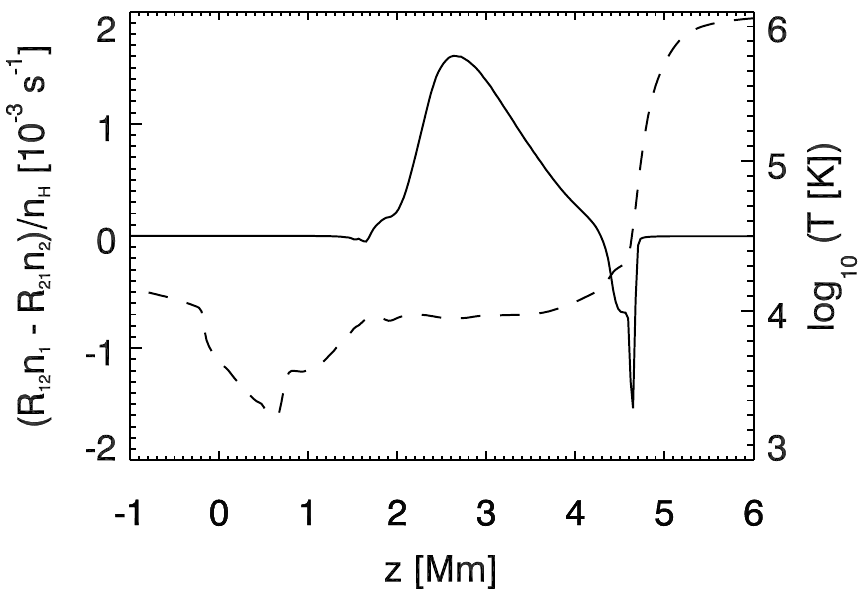}
 \caption{Net radiative rate per hydrogen atom in the Lyman-$\alpha$ transition (solid line,
                 scales to the left) and temperature (dashed line, scales to the right). Near the transition
                 region there is a net downward radiative rate in response to hydrogen in the ground state 
                 being excited by electron collisions followed by radiative de-excitation. Deeper down, in the chromosphere, there
                 is a net upward rate, where Lyman-$\alpha$ photons are absorbed. The data is from the
                 column indicated in the LYA-HION panels of Figure \ref{fig:zt_tempion}.}
 \label{fig:lya}
\end{figure}

All four simulations feature Lyman-$\alpha$ cooling, but only the LYA-HION-run and
the HELIUM-run include the Lyman-$\alpha$ heating and cooling self-consistently with
the rate equations. This self consistent treatment is the only thing separating the 
HION-run from the LYA-HION-run, so these two runs are used to identify the 
effects. We compare the temperatures in Figure~\ref{fig:tg_ions} (panels (b) and (c)), and the
temperature space-time cuts from  Figures~\ref{fig:zt_tempion} and 
\ref{fig:xt_temperature} (panels (b) and (c) in both figures). We do not see any systematic 
difference between the two runs. We do observe, 
however, when comparing the fraction of H as \ion{H}{2},  that including the 
Lyman-$\alpha$ heating leads to a more extended
region where the fraction of H as \ion{H}{2} is of order $\sim 10^{-1}$ 
(panels (r) and (s) from Figures~\ref{fig:tg_ions} and \ref{fig:zt_tempion}). This is expected for two
reasons, which we illustrate  in Figure \ref{fig:lya}. 
First, Lyman-$\alpha$ cooling in the transition region will contribute to a stronger 
net downward rate making it easier for atoms to remain neutral, even at high temperatures. 
Second, absorption deeper down, in colder regions, will contribute to a stronger net upward 
rate and more hydrogen occupying the first excited state. This results in more ions since 
photoionization in the Balmer continuum  is the most important ionization process for 
hydrogen in the chromosphere \citep{carlsson_stein2002}.

\subsection{Effects of Non-Equilibrium Helium Ionization}\label{section:helium}

\begin{figure*}
 \includegraphics[width=\textwidth]{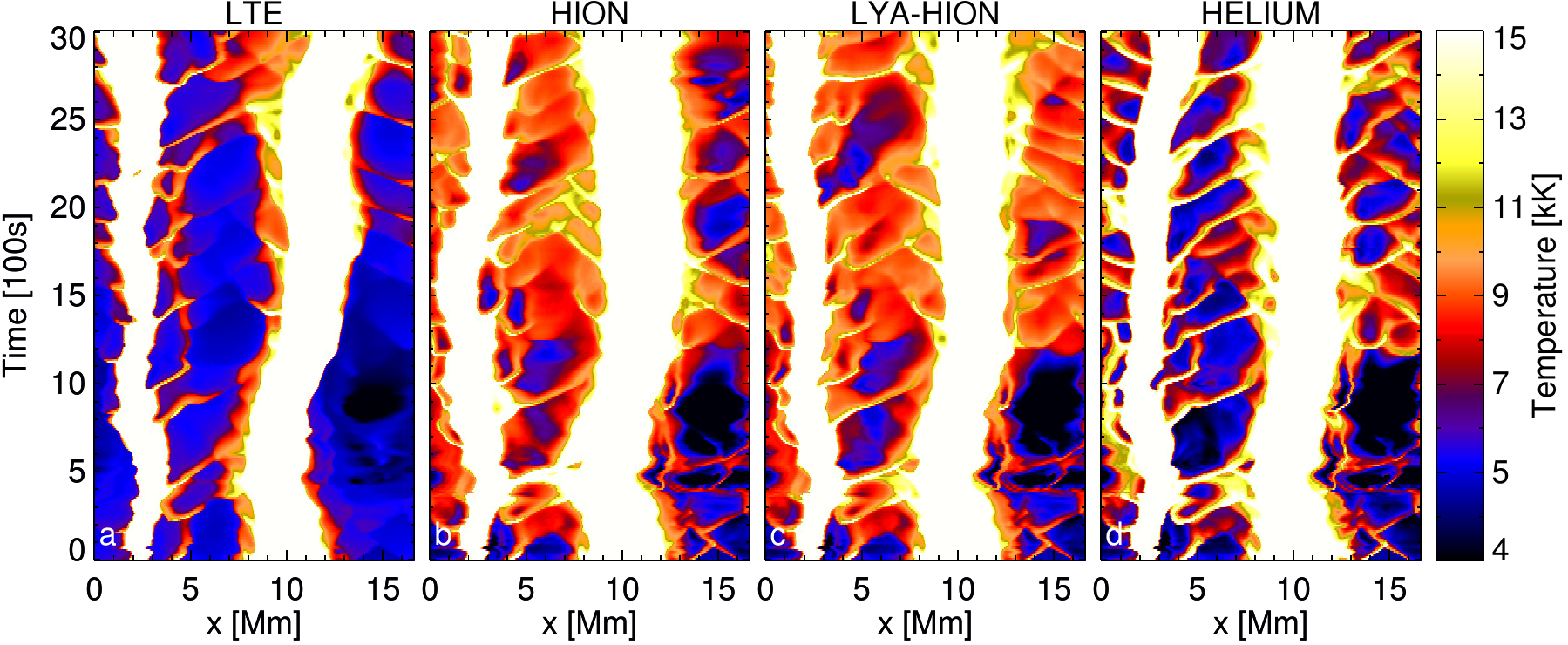}
   \caption{Temperature evolution along a horizontal cut ($z=2.66$Mm) in the four 
   simulation runs.  Waves propagating along the 
   magnetic field lines reveal themselves as relatively brighter inclined lines. 
   These waves stand out the most in the HELIUM run. Here the wave fronts are hotter
   than in any of the other runs, and the expanding gas between the waves is cooler. 
   The horizontal cut is indicated by the horizontal green line seen in all panels of Figure~
   \ref{fig:tg_ions}.}
 \label{fig:xt_temperature}
\end{figure*}

We now investigate the effects of non-equilibrium helium ionization and focus first on 
the two right columns of Figure~\ref{fig:tg_ions}, corresponding to the LYA-HION run and the
HELIUM-run. Panels (c) and (d) show the temperature. The difference is not dramatic, but there
is a tendency towards structural features in the chromosphere standing more out in the 
HELIUM-run. These structures are less pronounced in the HELIUM-run ion fractions. See for
instance the fraction of He as \ion{He}{2} in panel (l). It shows little or no correlation with the 
structures that can be seen in the chromosphere temperature in panel (d). 
Panel (k) displays this fraction in the LYA-HION-run and there is a solid correlation 
between it and the chromosphere structures showing in panel (c). 
The vertical and horizontal space-time temperature diagrams of 
Figures~\ref{fig:zt_tempion} and \ref{fig:xt_temperature} (panels (c) and (d) in both figures) indicate 
that the gas in the wave fronts is hotter, and gas between wave fronts is colder 
in the HELIUM-run than what they are in the LYA-HION-run. We compare the vertical 
space-time diagrams of the fraction of He as \ion{He}{2} 
(panels (k) and (l) of Figure \ref{fig:zt_tempion}) and see that these waves are essentially
not showing in the HELIUM-run, but they are in the LYA-HION-run. 

We explain the effect of non-equilibrium helium ionization in the same way we explain the
effect of non-equilibrium hydrogen ionization. The long ionization-recombination timescale
of helium prevents the ionization state to respond to waves and other perturbations. The increased
internal energy associated with the wave compression will, instead of ionizing the gas, lead to
increased temperatures. Conversely, the expanding gas between shocks cools off the material
instead of maintaining its temperature by releasing energy from recombining helium ions.

\begin{figure*}
 \includegraphics[width=\textwidth]{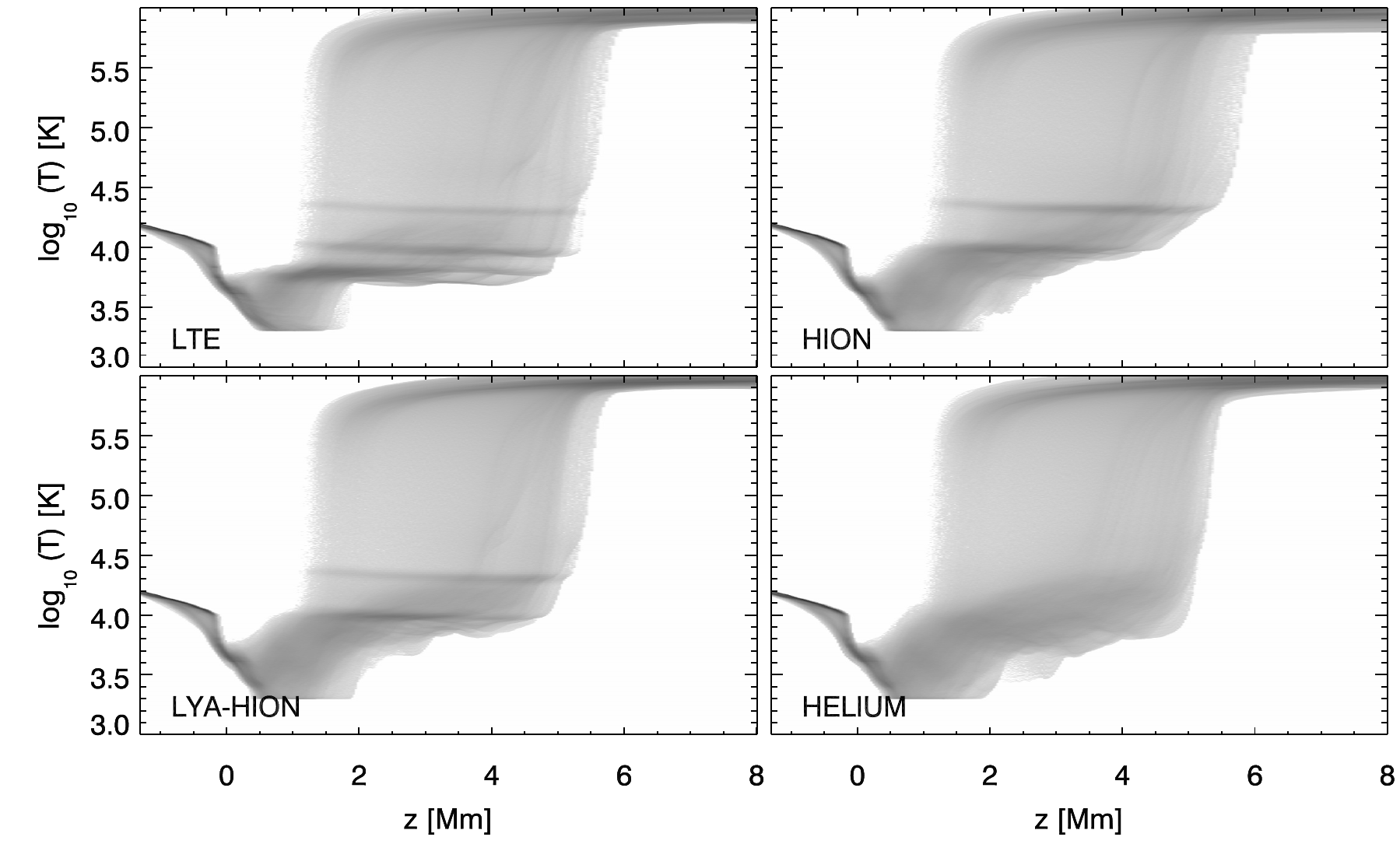}
 \caption{Joint Probability Density Function of 
 height and logarithmic temperature. The figure includes data from the
 time interval 1000-3000 seconds. The three horizontal plateaus (at 6, 10 and 22 kK) 
 in the LTE simulation indicate preferred temperatures when using the LTE 
 equation of state. These temperatures are associated with the LTE
 ionization of \ion{H}{1}, \ion{He}{1} and
 \ion{He}{2}. The plateaus vanish when we introduce non-equilibrium 
 hydrogen and helium ionization.}
 \label{fig:ztg_4histogram}
\end{figure*}

\subsection{Preferred Temperatures}\label{section:preferred_temperatures}

In the simulations where LTE ionization is used in the EOS, certain temperatures are more
frequently occurring than others. We can see this in the horizontal temperature space-time 
diagrams shown in Figure~\ref{fig:xt_temperature}. In the LTE-run (panel (a)) there 
are many points with a temperature around 6\,kK corresponding to blue color. In the two 
HION-runs there are many points with a temperature of about 10\,kK corresponding to 
orange/red color. Finally, in panel (d) we see that the temperature spans more of the 
color table range.

The preferred temperatures are also clearly visible as dark bands in Figure~\ref{fig:ztg_4histogram}, 
where we map out the occurrence rate of points on 
a $z-\log_{10}{T}$ grid. Using an LTE EOS clusters grid cells around these temperatures.
There is a cut in these panels at $\log_{10}{T} = 3.3$ (2\,kK). This is due to 
an artificial heating term that kicks in when the temperature drops below this value, 
effectively setting 2\,kK as a lower limit temperature.

If we heat a parcel of gas with LTE hydrogen ionization, the temperature will not rise above 
6\,kK before all of the hydrogen atoms are ionized. It is similar for LTE helium ionization. 
As we heat the parcel of gas, the temperature will not exceed 10\,kK before all of the
\ion{He}{1} atoms are  ionized. Heating it even more, 
eventually the temperature rises until we reach 22\,kK where all of the \ion{He}{2} ions  
are ionized before the temperature can increase to higher values. 
This happens because the Saha equation which governs the LTE ionization is 
particularly sensitive to temperature. Atoms or ions will  ionize over a 
small temperature range, while the corresponding range of internal energy is large.
In the chromosphere these small temperature ranges are centered at the preferred 
temperatures. There hydrogen and helium 
act as thermostats when their ionization state is 
described by LTE.

\subsection{EOS and Radiative Capability}
\begin{figure}
 \includegraphics[width=\columnwidth]{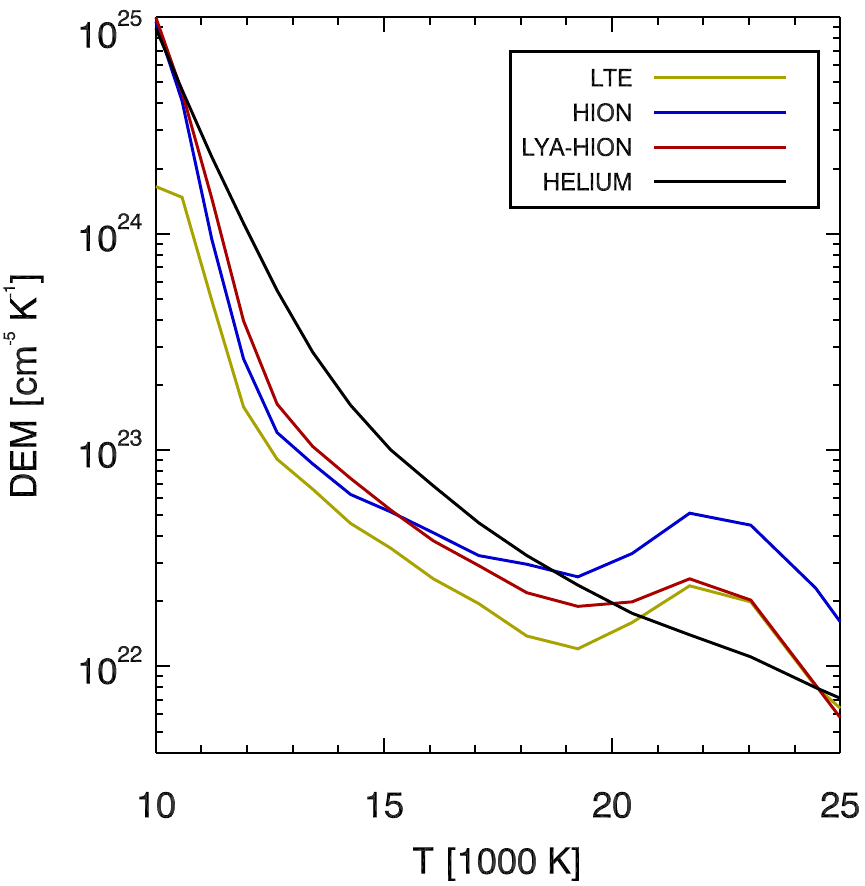}
 \caption{Differential emission measure averaged over the time
   interval 1000-3000 seconds. The HELIUM-run DEM does not have a bulge
   associated with the preferred temperature at $T=22$ kK, like the
   other three runs. It has a higher value than the other three
   runs in the temperature range 11-18 kK.}
 \label{fig:dem}
\end{figure}
\begin{figure}
 \includegraphics[width=\columnwidth]{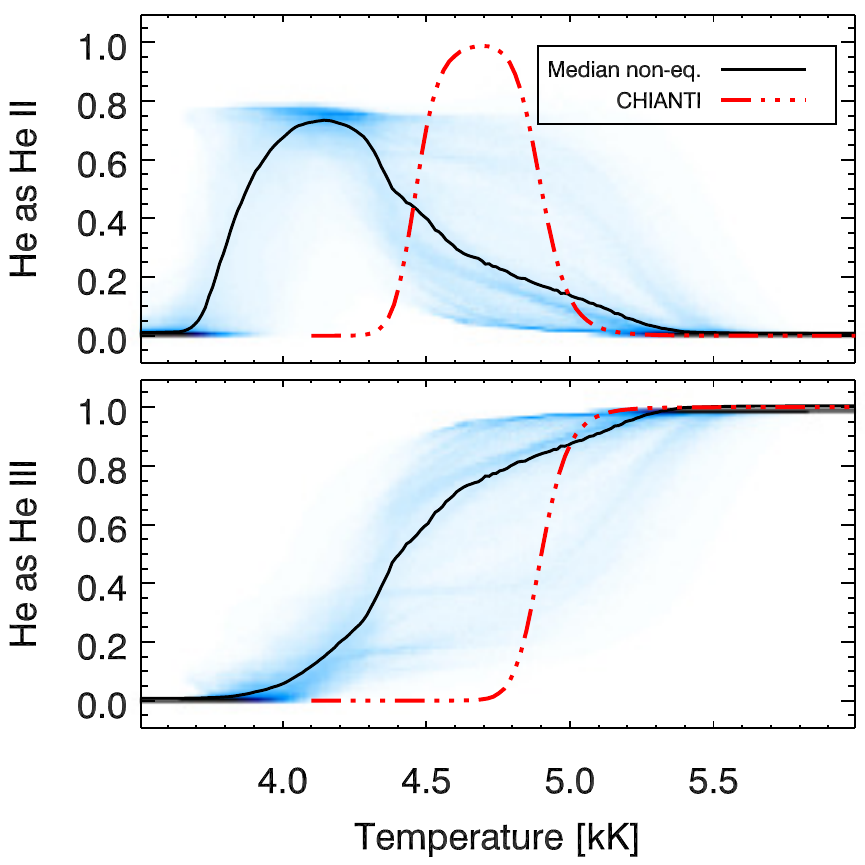}
 \caption{Occurrence of helium ion fractions as a function of temperature, \ion{He}{2} in
 top panel and \ion{He}{3} in bottom panel. Each column has been normalised to increase
 readability. Median values and CHIANTI values overplotted.}
 \label{fig:pdf_tg_ion}
\end{figure}

Using a non-equilibrium ionization EOS might change how a model atmosphere radiates.
We use the differential emission measure (DEM) to get a qualitative understanding
of this change. Assuming ionization equilibrium, the emergent intensity 
of a thin line from a column in our simulation can be expressed as,
\begin{eqnarray}
 I = \int G(T) \Phi(T) dT,\label{eq:dem}
\end{eqnarray}
where $G(T)$ is the line dependent contribution function determined by atomic data.
The DEM, $\Phi(T)$, is defined as $n_\mathrm{e} n_\mathrm{H} ~ dz/dT$. 
We compute this quantity on a
temperature grid ranging from 10-25\,kK. To get a statistically reasonable result, we average
over all columns of all the snapshots covering the timespan 1000-3000\,s. 

Figure \ref{fig:dem} shows the DEMs corresponding to the four simulation runs. 
The HELIUM-run DEM falls off smoother than the DEMs from the other runs.
At temperatures above 11\,kK it deviates from the two HION-run DEMs, whereas they are 
very similar at lower temperatures. In the temperature range 11-18\,kK the HELIUM-run 
DEM has a higher value by a factor of around 2 compared to the LYA-HION run. 
The bumps in the DEM curves at 22\,kK in all but the HELIUM-run are due to 
LTE ionization of helium and the resulting 22\,kK preferred temperature. Since the 
LTE-run has less material at temperatures between the two preferred temperatures 6\,kK and
10\,kK, its DEM in that temperature range is down by an order of $10^{-2}$ of the DEMs from the 
runs including non-equilbrium hydrogen ionization (not shown here). Although DEMs are used 
mostly for analysis concerning coronal lines formed at higher temperatures than the ones 
featured here, the quantity is correlated to the amount of mass and its ability to radiate at a 
given temperature - also for the low temperatures we are considering here. 

\subsection{Consequence for Modelling Helium Lines}
Using a DEM to model  helium resonance lines with Equation \ref{eq:dem} is not ideal for 
two reasons. First, the lines are not thin, and second, the assumption of ionization equilibrium 
is not valid. We elaborate on the latter in Figure \ref{fig:pdf_tg_ion}. Here we show the 
probability density functions of helium ions as a function of temperature, from the HELIUM-run. 
The time dependent fraction of He as \ion{He}{2} peaks at a temperature near 10\,kK - 
well below the corresponding ionization equilibrium value at around 50\,kK. This happens 
because the coronal radiation is photoionizing the "cold" and neutral helium in the upper 
chromosphere.

\subsubsection{Helium Ionization-Recombination Timescales}

\begin{figure}
 \includegraphics[width=\columnwidth]{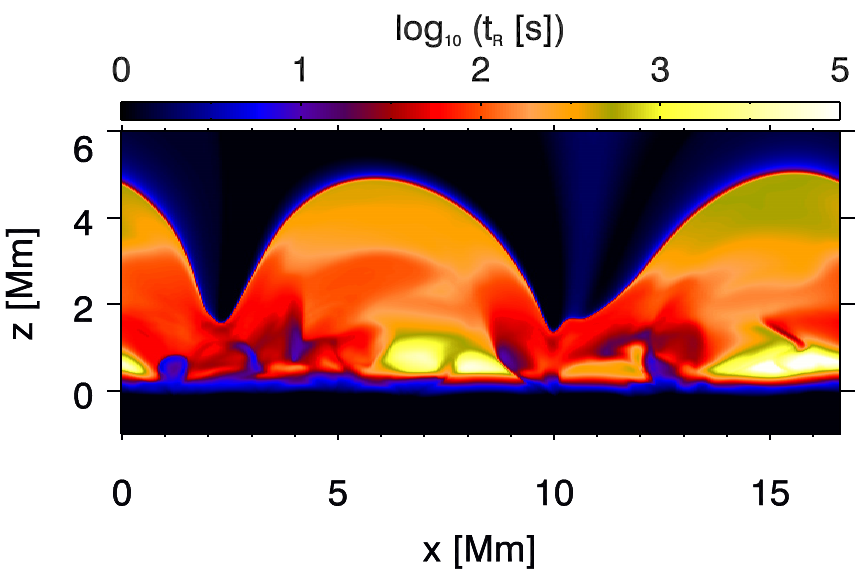}
 \caption{Ionization-recombination timescale of helium at $t=1400$ s.}
 \label{fig:timescale}
\end{figure}

We use the helium transition rates to compute $t_\mathrm{R}$, 
the ionization-recombination timescale \citep{judge2005}. This timescale is shown 
in Figure~\ref{fig:timescale} at $t=1400$\,s for the HELIUM-run (the same snapshot 
as in Figure~\ref{fig:tg_ions}).  Around the two network regions at $x=2$\,Mm 
and $x=10$\,Mm the timescales in the chromosphere are the shortest, ranging from $\sim10$\,s 
to $\sim 200$\,s. The transition region and corona above these regions are hot and dense, 
leading to high EUV emissivities ($\eta_{\mathrm{cor},j}$) and strong EUV heating of the 
chromosphere, i.e. large transition rates and short timescales. In the more elevated parts of the 
chromosphere residing under the arched structures, the timescale is an order of magnitude 
higher at  $\sim10^3$\,s. Here the gas has a low density and the incident radiation 
field is weaker than in the network regions, both of which lead to slow rates and 
long timescales.

\section{Conclusion}\label{section:conclusion}

We present a method for including non-equilibrium ionization of abundant
elements in the EOS of RMHD simulations. The method is implemented in the 
stellar atmosphere code Bifrost. To assess the effects of the dynamics of the chromosphere on the  ionization,
we present four 2D numerical simulations of the solar atmosphere. They feature
different setups for ionization in the EOS: it is either described by a set of Saha equations (LTE)
or by a set of rate equations for hydrogen and/or helium resulting in a non-equilibrium ionization 
state where possible long ionization-recombination timescales are taken into account.

We find that the use of a non-equilibrium ionization EOS affects the state of the 
chromosphere in the following way: there is a larger variation in the chromospheric
temperature which is caused by smaller variation in the ionization state than what 
is the case for LTE simulations. A slowly changing ionization state means
that the internal energy bound in ions is also slowly changing. Any short lived
chromospheric perturbation in the internal energy will therefore manifest itself as a perturbation in temperature. Both non-equilibrium hydrogen and helium ionization contribute
to this effect. The ionization state of helium impacts the upper chromosphere the most because 
hydrogen is already highly ionized there.

A side effect of using LTE ionization in the EOS is that grid cells that make up the chromosphere
and lower transition region, where the ionization state of the gas plays a decisive role in the 
energy balance, are likely to have one of three preferred temperatures, 6, 10 or 22\,kK,
associated with the ionization of \ion{H}{1}, \ion{He}{1} and \ion{He}{2} respectively.
This leads to chromospheric $(\rho, T)$ configurations that are not necessarily very physical, and 
that, in turn, may effect significantly how certain chromospheric 
lines form. We illustrate this with a DEM calculation that shows an increase by a factor two 
between 11~kK and 18~kK when including non-equilibrium helium ionization. This change in 
atmospheric structure might resolve part of the problem that  chromospheric spectral lines calculated 
from RMHD models typically show too little emission
\citep[\eg][]{leenaarts2013,2015ApJ...813...34L}.
Changes in the electron density and temperature might  affect the first ionization potential effect (FIP) observed in the corona
\citep[\eg][]{2015LRSP...12....2L}.
The non-equilibrium ionization of helium might influence the mechanisms that cause abundance differences of helium in the corona
\citep{2003ApJ...591.1257L},
and on the anomalous helium line intensities derived from DEM models
\citep{2015ApJ...803...66G}.

The timescale of helium ionization and recombination in the chromosphere span a range 
$10^1$ to $10^3$ seconds. The lower end of this range is consistent with the timescale we 
computed in \cite{golding2014}. There we found and gave examples of non-equilibrium effects 
on the \ion{He}{1} 10830 and \ion{He}{2} 304 lines. Since a timescale in the low end of the range 
is sufficient to cause non-equilibrium effects, these effects might be more pronounced in 
atmospheric regions  where the timescale is in the higher end of the range. 
We thus expect  a strong influence of non-equilibrium helium ionization on the formation of the \HeI~10830~\AA\ line, which is a popular diagnostics of chromospheric magnetic fields
\citep[\eg][]{2004A&A...414.1109L,2010ApJ...708.1579C,2013ApJ...768..111S}.

\edit{Our conlusions are based on a differential study using 2D simulations. 
Since the physical mechanisms do not change when going to 3D, we expect our conclusions to hold then as well.}

\edit{While our simulations with non-equilibrium hydrogen and helium ionization are an 
improvement over previous work, they have some limitations. Our treatment of the 
radiative transfer in the hydrogen Ly$\alpha$ and  \ion{He}{2} 304 lines is highly 
simplified. Also, ion-neutral interaction effects are not included, but they have been found 
to affect the thermodynamics of the chromosphere \citep{martinezsykora2012}.}

\edit{At the time of writing a 3D simulation including non-equilibrium ionization of hydrogen
and helium is underway. The natural next step would be the inclusion of ion-neutral
interactions.}

\acknowledgements

This research was supported by the Research Council of Norway through the 
grant ``Solar Atmospheric Modelling'' and through grants of computing time
from the Programme for supercomputing. The research leading to these results
has received funding from the European Research Council under the European Union's 
Seventh Framework Programme (FP7/2007-2013) / ERC Grant 
agreement nr. 291058 (CHROMPHYS).

\appendix
\section{Code Implementation of Non-Equilibrium Helium Ionization}
The variables included in experiments with non-equilibrium helium ionization are 
the temperature, $T$,  the electron density, $n_\mathrm{e}$, 
the atomic hydrogen population densities, $n_{i,\mathrm{H}}$, for five bound 
states ($i=1$ to $i=5$) and the continuum ($i=6$), the molecular hydrogen population 
density $n_{\mathrm{H2}}$, and, finally, the helium population densities $n_{i,\mathrm{He}}$ 
for the ground states of \ion{He}{1}, \ion{He}{2} and \ion{He}{3} ($i=1,2,3$).
This adds up to 12 variables, so we need to solve 12 equations. 
The equations are linearized and solved with the iterative Newton-Raphson 
scheme. The first 9 equations are the 
internal energy conservation equation (Eq. \ref{eq:energyeq}), charge conservation
equation (Eq. \ref{eq:chargeeq}), a hydrogen atom conservation equation,
and six rate equations, one for molecular hydrogen and five for atomic hydrogen.
For the three last equations we use two rate equations for atomic helium 
and one helium atom conservation equation. 

The helium and hydrogen atom conservation equations have identical forms. For brevity 
we let $n_i$ refer to either $n_{i,\mathrm{H}}$ or $n_{i,\mathrm{He}}$ and express the 
atom conservation equations:
\begin{eqnarray}
  F_\mathrm{cons} &=& \frac{\sum_{i=1}^{N} 
              n_i}{n_{\mathrm{tot}}} -1=0,
       \label{eq:hecons}
\end{eqnarray}
where $n_{\mathrm{tot}}$ is the total number density of either hydrogen or helium, 
which is proportional to the mass density with an abundance dependent coefficient, and
$N$ is the number of atomic states. For hydrogen $N$ is 6 and for helium it is 3.
The atom conservation equations only depend on the number density and the derivative
is
\begin{eqnarray}
\frac{\partial F_{\mathrm{cons}}}{\partial n_{i}} &=& \frac{1}{n_{\mathrm{tot}}}
\end{eqnarray}

The atomic rate equations of hydrogen and helium also have identical forms. We define
\begin{eqnarray}
P_{ii}=-\sum_{j,j\ne i}^N P_{ij},
\end{eqnarray}
where $P_{ij}$ is the sum of the collisional and radiative rate coefficient
for the transition from atomic state $i$ to atomic state $j$: $P_{ij} = C_{ij} + R_{ij}$.
The atomic rate equations then take the form
\begin{eqnarray}
 F_{i,\mathrm{rate}} &=& \frac{\Delta t}{n_{i}^{\mathrm{old}}} 
        \sum_{j= 1}^{N} n_{j} P_{ji}
       -  \frac{n_{i}}{n_{i}^{\mathrm{old}}} +1=0.
       \label{eq:disc_rateeq}
\end{eqnarray}
These equations depend on electron density, temperature and number densities.
The derivatives are:
\begin{eqnarray}
 \frac{\partial F_{i,\mathrm{rate}}}{\partial n_{\mathrm{e}}} &=& 
                    \frac{\Delta t}{n_{i}^{\mathrm{old}}}
                    \sum_{j=1}^N n_{j} \frac{\partial P_{ji}}{\partial n_{\mathrm{e}}} \\
 \frac{\partial F_{i,\mathrm{rate}}}{\partial T} &=& 
                    \frac{\Delta t}{n_{i}^{\mathrm{old}}}
                    \sum_{j=1}^N n_{j} \frac{\partial P_{ji}}{\partial T} \\
 \frac{\partial F_{i,\mathrm{rate}}}{\partial n_{j}} &=& 
                    \frac{\Delta t}{n_{i}^{\mathrm{old}}}
                      \ P_{ji} , \ \ \mathrm{where} \ j\ne i \\
\frac{\partial F_{i,\mathrm{rate}}}{\partial n_{i}} &=& 
                    \frac{\Delta t}{n_{i}^{\mathrm{old}}}
                       P_{ii} - \frac{1}{n_{i}^{\mathrm{old}}} 
\end{eqnarray}
In the old non-equilibrium hydrogen description (the HION-run), all Lyman transitions
are assumed to be in detailed radiative balance, i.e. the radiative rate coefficients
to or from the ground state of hydrogen ($i=1$) are \edit{all set to zero}. We have derived approximate
expressions for the Lyman-$\alpha$ and Lyman continuum radiative rate coefficients,
$R_{12}$, $R_{21}$, $R_{16}$ and $R_{61}$ given in Equations \ref{eq:rijlya}, \ref{eq:rjilya},
\ref{eq:rij} and \ref{eq:rji}, respectively. Of these rate coefficients, only $R_{61}$ has 
non-zero derivatives. These are given in the next section.
For all other derivatives we refer the reader to the Appendix of \cite{leenaarts2007}. 
For the rate equation for molecular hydrogen and its derivatives we refer the reader 
to Appendix B of \cite{gudiksen2011}.

\section{Transition Rate Coefficients}
The total transition rate coefficient needed for the rate equations, $P_{ij}$, is the sum 
of a collisional part, $C_{ij}$ and
a radiative part $R_{ij}$. In this section we give a derivation of the radiative
rate coefficients and provide expressions for the collisional rate coefficients. 

\subsection{Derivation of Photoionization Rate Coefficients}\label{section:rate_coeff}
The general expression for the photoionization rate coefficient \citep[\eg][]{mihalas1978} is 
given by
\begin{eqnarray}
  R^{\mathrm{up}}_{\mathrm{x}} = 4\pi \int_{\nu_0}^{\infty} 
                   \frac{\sigma_{\mathrm{x},\nu}}{h\nu} J_\nu \ d\nu,
\end{eqnarray}
where $\sigma_{\mathrm{x},\nu}$ is the frequency dependent photoionization 
cross section and $J_\nu$ is frequency dependent mean intensity, $h$ is Planck's constant,
 and $\nu_0$ corresponds to the difference in energy between the ground and continuum states. 
The subscript x denotes the transition under consideration. 
We have adopted a bin formulation. In each bin the photoionization cross 
section is constant and determined by Equation \ref{eq:photoxbin}. The
mean intensity is also constant for each bin, $J_j$. $\nu_{j,0}$ and $\nu_{j,1}$ denotes the lower and upper 
frequency boundaries of the $j$th bin. We can then express the photoionzation
rate coefficient as a weighted sum over the binned mean intensity,
\begin{eqnarray}
  R^{\mathrm{up}}_{\mathrm{x}} &=& 
                    \sum_{j=1}^{N_\mathrm{bin}} W_{\mathrm{x},j} J_j,
\end{eqnarray}
where the weights are defined,
\begin{eqnarray}
  W_{\mathrm{x},j} = \frac{4\pi \sigma_{\mathrm{x},j}}{h} \log{ 
  \left( \frac{\nu_{j,1}}{\nu_{j,0}}\right)}.
\end{eqnarray}
The radiative recombination coefficient, 
ignoring stimulated emission, is expressed
\begin{eqnarray}
  R^{\mathrm{down}}_{\mathrm{x}} = 4\pi \left[ \frac{n_{\mathrm{g},\mathrm{x}}}
                  {n_{\mathrm{c},\mathrm{x}}}\right]_{\mathrm{LTE}} \int_{\nu_0}^{\infty}
                  \frac{\sigma_{\mathrm{x},\nu}}{h\nu}B_{\nu}^{\mathrm{sp}} \ d\nu,
\end{eqnarray}
where the Planck function neglecting stimulated emission is
\begin{eqnarray}
  B_\nu^{\mathrm{sp}} = \frac{2h\nu^3}{c^2} e^{-h\nu/k_{\mathrm{B}}T}.
\end{eqnarray}
Here $c$ is the speed of light and $k_\mathrm{B}$ is Boltzmann's constant.
$\left[n_{\mathrm{g,x}} / n_{\mathrm{c,x}} \right]_{\mathrm{LTE}}$ is the LTE 
ground state to continuum ratio. Given the statistical weights,
$g_{\mathrm{g,x}}$ and $g_{\mathrm{c,x}}$, the electron mass, $m_\mathrm{e}$,
and the difference in energy between the ground and continuum state, $\chi_\mathrm{x}$,
the LTE ratio is expressed: 
\begin{eqnarray}
   \left[   \frac{n_{\mathrm{g},\mathrm{x}}}  {n_{\mathrm{c},\mathrm{x}}}\right]_{\mathrm{LTE}} =
       \frac{n_\mathrm{e}}{2} \frac{g_{\mathrm{g,x}}}{g_{\mathrm{c,x}}} 
             \left( \frac{h^2}{2\pi m_e T} \right)^{3/2} 
               e^{\chi_\mathrm{x}/k_\mathrm{B}T}
\end{eqnarray}
In the bin formulation we use the binned photoionization cross 
section, $\sigma_{\mathrm{x},j}$ and a bin integrated 
$B^{\mathrm{sp}}_\nu$. The radiative recombination coefficient then becomes
a weighted sum, much like the photoionization coefficient,
\begin{eqnarray}
 R_{\mathrm{x}}^{\mathrm{down}} &=& \left[   \frac{n_{\mathrm{g},\mathrm{x}}}
                  {n_{\mathrm{c},\mathrm{x}}}\right]_{\mathrm{LTE}} \sum_{j=1}^{N_{\mathrm{bin}}}
                  W_{\mathrm{x},j} B_j,
\end{eqnarray}
where 
\begin{eqnarray}
  B_j = \frac{1}{\nu_{j,1} - \nu_{j,0}} \int_{\nu_{j,0}}^{\nu_{j,1}} B_\nu^{\mathrm{sp}} \ d\nu.     
  \label{eq:binplanck}
\end{eqnarray}

All derivatives of  $R_\mathrm{x}^\mathrm{up}$ are zero. $R_\mathrm{x}^{\mathrm{down}}$
depends on the electron density and the temperature. The derivatives are:
\begin{eqnarray}
  \frac{\partial R_\mathrm{x}^{\mathrm{down}}}{\partial n_\mathrm{e}} &=& 
            \frac{R_\mathrm{x}^{\mathrm{down}}}{n_{\mathrm{e}}} \\ 
  \frac{\partial R_\mathrm{x}^\mathrm{down}}{\partial T} &=& 
            R_\mathrm{x}^\mathrm{down} \left( -\frac{3}{2T} - 
                                   \frac{\chi_\mathrm{x}}{k_{\mathrm{B}}T^2} \right)   \nonumber \\
       && \ \ \ +\left[ \frac{n_\mathrm{g,x}}{n_\mathrm{c,x}} \right]_{\mathrm{LTE}} 
                   \sum_{j=1}^{N_\mathrm{bin}} W_{\mathrm{x},j} \frac{d B_j}{dT},
\end{eqnarray}
where
\begin{eqnarray}
  \frac{d B_j}{dT}  &=& \frac{1}{\nu_{j,1} - \nu_{j,0}} \int_{\nu_{j,0}}^{\nu_{j,1}}
               \left( \frac{h\nu}{kT^2} \right) B_{\nu}^{\mathrm{sp}} \ d\nu.
\end{eqnarray}
In the code we interpolate in pre computed tables of $\ln{(B_j)}$ and $\ln{(dB_j/dT)}$
as functions of $\ln{(T)}$.

\subsection{Collisional ionization/recombination rate coefficients}\label{section:coll_coeff}
The collisional ionization and recombination 
rate coefficients are 
\begin{eqnarray}
  C_{ij} &=& n_\mathrm{e}q_{ij}(T) \\
  C_{ji} &=& \left[ \frac{n_{i,\mathrm{He}}}{n_{j,\mathrm{He}}} \right]_{\mathrm{LTE}} C_{ij},
\end{eqnarray}
where $q_{ij}(T)$ is a temperature dependent function described in \cite{ar1985}. The 
derivatives are
\begin{eqnarray}
  \frac{\partial C_{ij}}{\partial n_\mathrm{e}} &=& \frac{C_{ij}}{n_\mathrm{e}} \\
  \frac{\partial C_{ij}}{\partial T} &=& n_\mathrm{e} \frac{dq_{ij}}{dT} \\
  \frac{\partial C_{ji}}{\partial n_\mathrm{e}} &=& 2\frac{C_{ji}}{n_\mathrm{e}} \\
  \frac{\partial C_{ji}}{\partial T} &=& C_{ji} 
                              \left(-\frac{3}{2T} - \frac{\chi_\mathrm{x}}{k_\mathrm{B} T^2} \right) \nonumber \\
         && \ \ \ + \left[ \frac{n_\mathrm{g,x}}{n_\mathrm{c,x}} \right]_{\mathrm{LTE}} 
                         \frac{\partial C_{ij}}{\partial T}
\end{eqnarray}
In the code we interpolate pre computed tables containing $\ln{(q_{ij})}$ and $\ln{(dq_{ij}/dT)}$ 
as functions of $\ln{(T)}$. 

\bibliographystyle{apj}
\bibliography{references}

\end{document}